\documentclass[letter]{aa} 

\usepackage{graphicx,hyperref}
\usepackage{txfonts}
\usepackage{natbib}
\begin{document}

   \title{Possible astrometric discovery of a substellar companion to the closest binary brown dwarf system WISE~J104915.57--531906.1\thanks{Based on data obtained with the ESO Very Large Telescope under programme 291.C-5004.}
   }
\titlerunning{Astrometric companion to WISE~J104915.57--531906.1}
\authorrunning{H.M.J.~Boffin et al.}
   \author{H.M.J. Boffin
          \inst{1}
          \and
          D.~Pourbaix\inst{2}\fnmsep\thanks{Senior Research Associate, F.R.S.-FNRS,
  Belgium}
          \and
          K. Mu\v zi\' c
          \inst{1}
          \and
          V.D. Ivanov
          \inst{1}
\and
          R. Kurtev
          \inst{3}
                  \and
          Y. Beletsky
          \inst{4}
             \and\\
          A. Mehner
          \inst{1}
            \and
         J.P. Berger
          \inst{5}
  \and
          J.H. Girard
          \inst{1}
  \and
         D. Mawet
                   \inst{1}
          }

   \institute{ESO, Av.~Alonso de Cordova 3107, Casilla 19001, Santiago 19, Chile\\
              \email{hboffin@eso.org}
         \and		
Institut d'Astronomie et d'Astrophysique, Universit\'e Libre de Bruxelles
  (ULB), Belgium      
         \and
             Departamento de F\'\i sica y Astronom\'\i a, Universidad de Valparaiso, Av. Gran Breta\~na 1111, Playa Ancha, 5030, Casilla, Chile
         \and
         	Las Campanas Observatory, Carnegie Institution of Washington, Colina el Pino, Casilla 601 La Serena, Chile
          \and
	ESO, Karl-Schwarzschild-str. 2, 85478 Garching, Germany
   }

   \date{Received November XX, 2013; accepted XXX XX, 2013}

 
  \abstract
   {
Using FORS2 on the Very Large Telescope, we have astrometrically monitored over a period of two months the two components of the brown dwarf system \object{WISE J104915.57-531906.1}, the closest one to the Sun. Our astrometric measurements -- with a relative precision at the milli-arcsecond scale -- allow us to detect the orbital motion and derive more precisely the parallax of the system, leading to a distance of 2.020$\pm$0.019 pc. The relative orbital motion of the two objects is found to be perturbed, which leads us to suspect  the presence of a substellar companion around one of the two components. We also perform $VRIz$  photometry of both components and compare with models. We confirm the flux reversal of the T dwarf.}


   \maketitle
%

\section{Introduction}

Very recently, \citet{Luhman-2013:a} identified WISE J104915.57--531906.1 (hereafter Luhman 16AB, to follow the nomenclature of \citealt{2013arXiv1307.6916B}), as a very red binary star with an incredibly high proper motion of $\sim 3\arcsec$/yr and a distance
of $\sim2$ pc, making it the third closest system to
the Sun, after $\alpha$~Cen and Barnard's star. The two objects in Luhman 16AB are separated by 1.5\arcsec, with the primary being a brown dwarf (BD) of spectral type L8.  

In \citet{2013ApJ...770..124K}, we reported comprehensive follow-up observations of this newly detected system, confirming the spectral types of both BDs (L8$\pm$1, T1$\pm$2) and, based on the small relative radial velocity of the two components, confirmed that they form a gravitationally bound system.  Our $JHK_S$ photometry yields colours consistent with the spectroscopically derived spectral types, while a comparison of the apparent magnitudes with models leads to a distance of $\sim2.25$ pc, in agreement with the parallax of \citet{Luhman-2013:a}. The available kinematic and photometric information exclude the possibility that the object may belong to any of the known nearby young moving groups nor associations. For the given spectral types and absolute magnitudes, 1 Gyr theoretical models -- DUSTY \citep{DUSTY} and BT-Settl \citep{2011ASPC..448...91A}  --
predict masses of 0.04--0.05~M$_\odot$ for the primary, and 0.03--0.05~M$_\odot$ for the secondary. The objects remain in the sub-stellar regime even if they are 10 Gyr old. 

\citet[][see also \citealt{2013arXiv1310.5144B}]{2013A&A...555L...5G} reported complex quasi-periodic (4.87 h) near-infrared photometric variability at  $\sim$0.1~mag level, which they associate with the rotation of the secondary. They did not detect any transit during their 12 night monitoring.    \citet{2013ApJ...772..129B} also reported resolved near-infrared spectroscopy and photometry of Luhman 16AB, which reveal strong H$_2$O and CO absorption features in the
spectra of both components, the secondary also showing weak CH$_4$ absorption. They found the system to exhibit a ``flux reversal'', that is, the T dwarf secondary appears brighter in the $0.9-1.3~\mu$m wavelength range but is fainter at shorter wavelengths. 

In this paper, we present further imaging observations of the individual components of Luhman 16AB, with the aim to start a long term astrometric campaign, that should lead to a more definitive parallax and the orbit of the system. 
Based on GMOS observations and  additional archival data from DSS, DENIS, and 2MASS, \citet{Luhman-2013:a} ended up with a set of six positions over 23 years yielding a parallax ($\varpi$) of $496\pm37$ mas.  He also obtained $-2\,759\pm6$ and $354\pm6$ mas\,yr$^{-1}$, respectively, for the proper motion $\mu_{\alpha*}$ ($=\mu_{\alpha}\cos\delta$) and $\mu_{\delta}$.  In a note on {\tt astro-ph}, \cite{Mamajek-2013:a} compiled some older positions from the ESO Schmidt red and GSC 2.3 catalogues. Using these additional points, we obtain a revised parallax of $\varpi=514\pm26$ mas.



\section{Observations and data reduction}

Over a period of two months starting mid-April 2013, Luhman 16AB was observed with the FORS2 multi-purpose optical instrument \citep{FORS2}  attached to Unit 1 of the Very Large Telescope. The high-resolution collimator was used with 2 $\times$ 2 binning, resulting in a pixel scale of 0.125\arcsec\, and a field of view of 4.1\arcmin. The MIT CCD was used, which is comprised of two chips, separated by a gap of  10.8\arcsec. Luhman 16AB lies in the upper chip (CHIP1) and in this study we have only used stars on this chip to avoid introducing the additional misalignment between the two chips. 

Observations were made using the $I$-band filter (I\_BESS+77) on 12 epochs over the period from  14 April to 22 June 2013, with each epoch generally separated by 5 or 6 days. For each epoch, at least 21 images were taken (although not all were used, see below). The exposure time ranged from 15 to 60 seconds, and the seeing was always better than 1.2\arcsec, ensuring that the two components of the binary system were always well separated. Observations were done only when the object was at airmass below $\sim$1.2. A detailed log of the observations is provided in the online Table~\ref{tab:ObsLog}, while a typical image is shown in online Fig.~\ref{Fig:Obs}.

The individual images of a single night were stacked together by minimising the scatter in $\Delta\alpha*= \Delta\alpha \cos \delta$ and $\Delta\delta$ for every source, where as usual, $\alpha$ and $\delta$ are the right ascension and the declination, respectively.  
The sample of images was limited to those where the two components of Luhman 16AB were resolved and measured. On each night, a reference image was adopted and the median of the shift derived for all sources but Luhman 16AB was determined.  The resulting uncertainty on the positions, between 2 and 10 mas (see the values $\sigma_{\alpha*}$ and $\sigma_{\delta}$ in the online Table~\ref{tab:ObsLog}), was computed as the standard deviation of the 5$^{\rm th}$--95$^{\rm th}$ percentiles of the stacked positions.  The same procedure was then applied to every night relative to the first one.  The ICRS coordinates were finally restored by measuring the position of one star in one of these images.

The accuracy of such a stacking procedure heavily relies upon a key assumption: no other point source on the field of view is moving over the two month observation campaign.  Point source here means a single star with a noticeable parallax and/or proper motion or a binary (unresolved or with only one component visible) with a significant orbital motion.  At the 5-mas level, no disturbing point source is present in the field of view, thus making the stacked image pretty robust.  

In addition to our astrometric monitoring in the $I$-band, we have also obtained on 16 June 2013, images of Luhman 16AB with FORS2 in the $V$, $R$ and $z$ bands, with the v\_HIGH, R\_SPECIAL and z\_GUNN filters, and with exposure times of 480, 120 and 5 seconds, respectively. The resulting images are also shown in online Fig.~\ref{Fig:Obs}.

\section{Astrometric models}
Two sets of observations are available: the absolute positions of the photocentre over 23 years from archival data and the absolute positions of both components over our two month monitoring.  Even if the latter are much more precise, a campaign of two months is too short to derive either the parallax or the proper motion.  They can nevertheless be combined with the older data to improve the precision of the fundamental astrometric parameters which constrain the size and orientation of the orbit.  On the other hand, the positions of the two components can be changed into relative positions of one component with respect to the other, thus removing the effect of the parallax and proper motion and keeping the orbital parameters only.

\begin{figure*}[tbp]
\includegraphics[width=6cm]{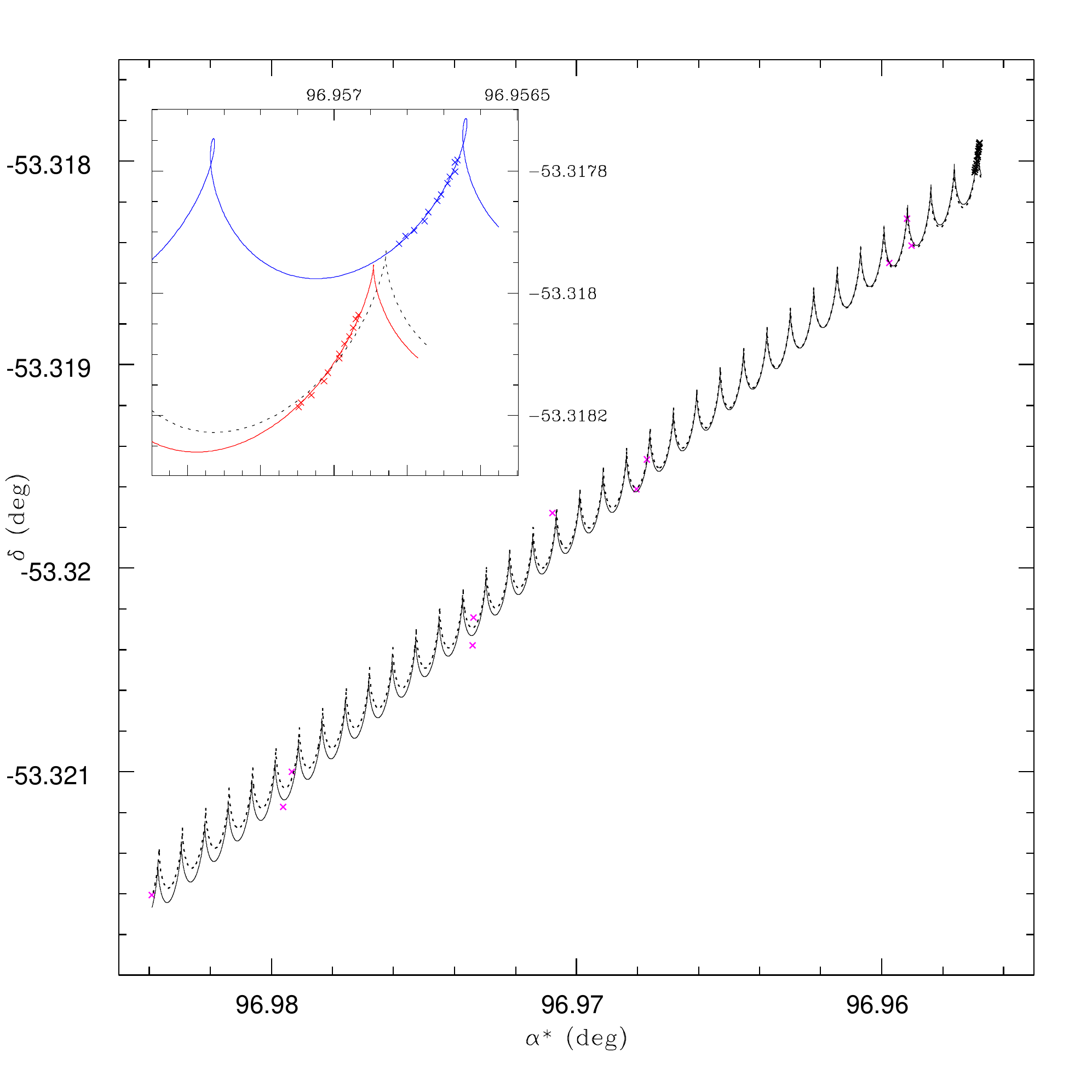}
\includegraphics[width=6cm]{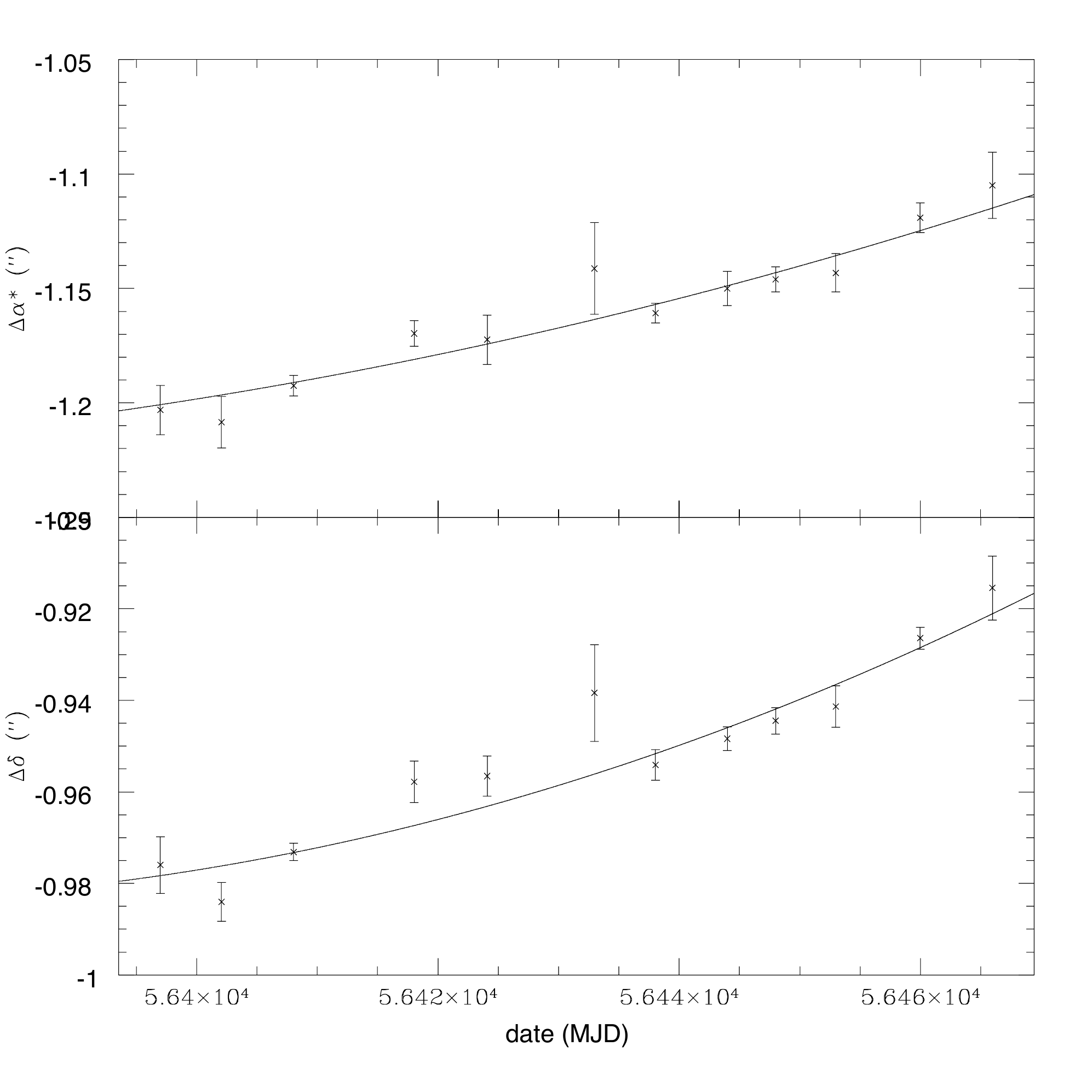}
\includegraphics[width=6cm]{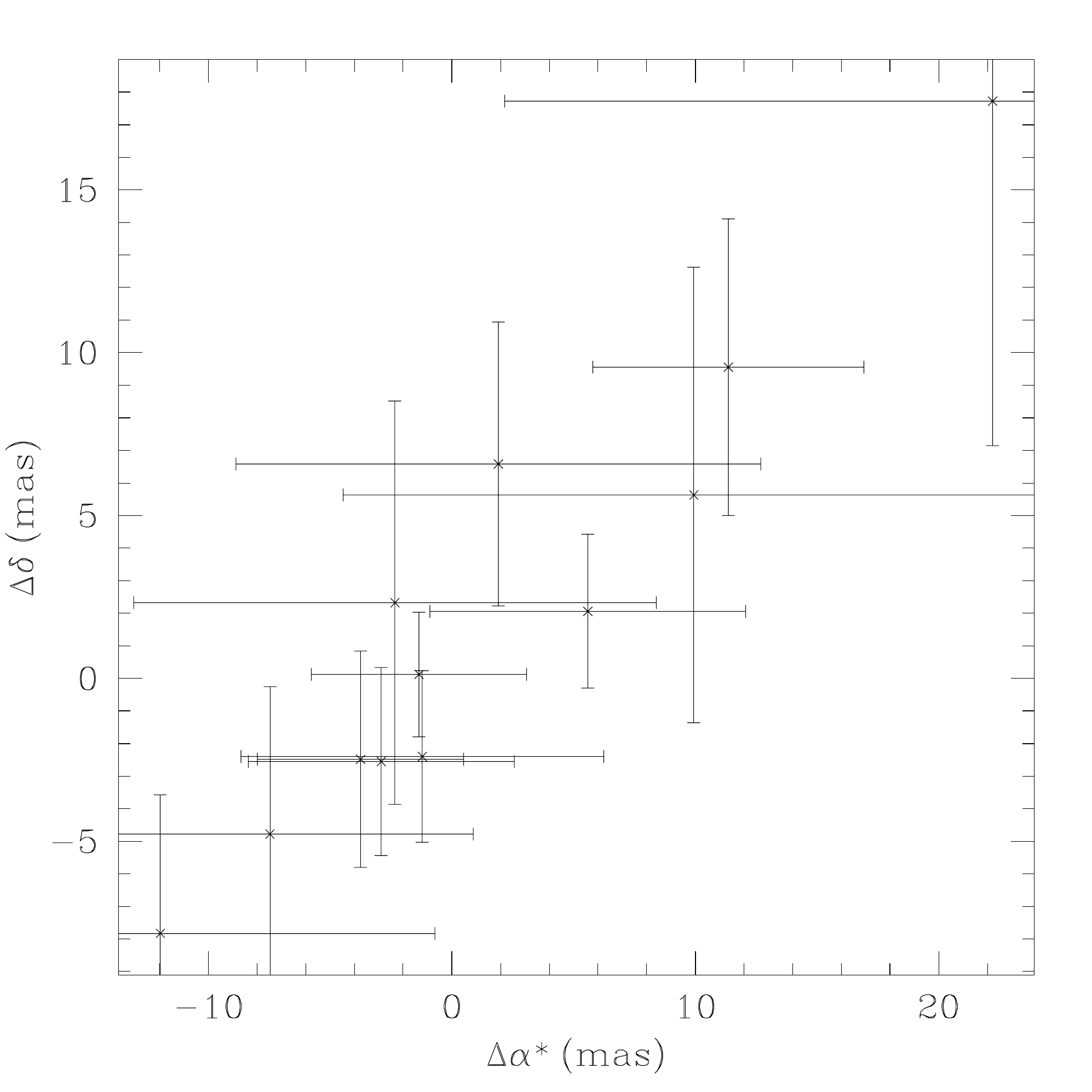}
\caption[]{\label{fig:AbsPlot}(left) Old and new observations adjusted together.  The dotted line represent the solution based upon \citet{Luhman-2013:a} and \citet{Mamajek-2013:a}.  The black continuous line represent the motion of the barycentre from our solution.  The inset zooms onto  the region covered by the FORS2 observations.  The red (resp. blue) line represents the motion of the L (resp. T) dwarf.
(middle) Relative positions obtained with FORS2 during our 2-month monitoring campaign and the best parabolic fits. 
(right) Residuals of the FORS2 data based on the parabolic fit.
}
\end{figure*}

\subsection{Two-body orbital model}\label{sect:twobody}
The motion of a resolved binary is fully described by the motion of its barycentre (position, parallax, and proper motion) and the orbital motion of each component around it \citep{PrDoSt}.  The two orbits only differ by their size and the argument of the periastron, $180^{\circ}$ apart.  Thirteen parameters are thus required.  However, a careful selection of these parameters can make the fitting procedure much more straightforward.  For instance, adopting the Thiele-Innes parameters for, say, the primary make nine parameters appear linearly, instead of only five with the Campbell orbital elements \citep[e.g.][]{2003A&A...398.1163P}.  With the orbit of the primary setting up the unit, the orbit of the secondary is just its scaled image, the scaling factor ($\rho$) lying between 0 and $+\infty$.  In practice, a grid search was used for the remaining four nonlinear parameters.

Even if every step leading to the stacked image was performed very thoroughly, the transformation of the relative positions into the absolute ones relies upon one direct measurement only, thus making it the single point of failure.  In order to restore some robustness, two linear parameters ($o_{\alpha*}$ and $o_{\delta}$) are therefore added to cope with some misalignment between the old and new dataset.
The resulting model is therefore given by:
\begin{eqnarray*}
\xi&=&\xi_0+o_{\alpha*}+\varpi f_a+\mu_{\alpha*}(t-2000.0)+BX+GY\\
\eta&=&\eta_0+o_{\delta}+\varpi f_d+\mu_{\delta}(t-2000.0)+AX+FY
\end{eqnarray*}
for the primary,
\begin{eqnarray*}
\xi&=&\xi_0+o_{\alpha*}+\varpi f_a+\mu_{\alpha*}(t-2000.0)-B\rho X-G\rho Y\\
\eta&=&\eta_0+o_{\delta}+\varpi f_d+\mu_{\delta}(t-2000.0)-A\rho X-F\rho Y
\end{eqnarray*}
for the secondary, and
\begin{eqnarray*}
\xi&=&\xi_0+\varpi f_a+\mu_{\alpha*}(t-2000.0)\\
\eta&=&\eta_0+\varpi f_d+\mu_{\delta}(t-2000.0)
\end{eqnarray*}
 for the barycentre. $f_a$ and $f_d$ denote the parallactic factors \citep{FuAs}, $\xi=\alpha\cos\delta$, and $\eta=\delta$. \citet{Luhman-2013:a} assumed that, on the old plates, the measured photocentre matches the epoch barycentre.  As the exact orbit is not available yet, one cannot say how wrong that assumption is.  During our observation campaign, the offset of the photocentre with respect to the barycentre would range from 40 mas in $I$ to 100 mas in $V$.  The resulting least-squares fit (with a grid for the few non-linear parameters) is illustrated on Fig.~\ref{fig:AbsPlot}.

\subsection{Quadratic approximation}\label{sect:quadapprox}
Even if two components exhibit a relative motion which prevents them from being fully described by two single star solutions (sharing a common parallax and proper motion), that relative motion is so small that using the full orbital model from the previous section to describe it is simply an {\it overkill}
as it offers too much freedom with respect to the number of data points available to constrain the parameters of the model.  For an orbital period long enough, over an observation campaign of two months, both $\Delta\xi$ and $\Delta\eta$ can be modeled with an arc of parabola without biasing any conclusion with respect to the true orbital Keplerian model.  
Over the full orbital model, this quadratic one also offers the advantage of being linear, thus yielding a straightforward unique least-squares solution.

The validity of such a simplification was assessed through a large scale Monte-Carlo simulation. Five million orbits were generated (random orientation, period normally distributed around 20 years).  For all of them, twelve positions were derived for the same dates as the actual observations and a likely noise added.  At the $3\sigma$-level, no quadratic fit was distinguishable from the genuine orbital one.

\subsection{Results}
The results of the least square fit of the absolute positions 
 (i.e. model from Sec.~\ref{sect:twobody})
 from \citet{Luhman-2013:a}, \citet{Mamajek-2013:a}, and FORS2 are given in Table~\ref{tab:results} and plotted on Fig.~\ref{fig:AbsPlot}a.  The contribution of the primary to the $\chi^2$ is 20\% larger than the secondary, thus suggesting that a companion might be present around the former, making it oscillate. 

\begin{table}[tbp]
\centering
\caption[]{\label{tab:results}Astrometric results based on the old and new absolute positions and those obtained by Luhman (2013).}
\begin{tabular}{lrr}
\hline
\hline
Parameter & This work & Luhman \\
& & (2013)~~\\
\hline
$\varpi$ (mas) & $495\pm4.6$& $496\pm37$\\
$\mu_{\alpha*}$ (mas\,yr$^{-1}$) & $-2\,763\pm2.7$&$-2\, 759\pm 6$\\
$\mu_{\delta}$ (mas\,yr$^{-1}$) & $363\pm4.1$&$354\pm6$\\
\hline
\end{tabular}
\end{table}

The parabolic least-square fit of the relative positions is more difficult to assess. Both $\Delta\alpha*$ and $\Delta\delta$ are fitted with distinct parabolae (Fig.~\ref{fig:AbsPlot}b).  The larger $\chi^2$ is then used to derive the probability of rejecting the parabola (i.e. a two-body model) by accident even though it holds.  Whereas, for a single parabola, such a probability is tabulated and directly available, we relied upon extensive Monte-Carlo simulations to quantify the effect of
taking the larger of the two $\chi^2$ rather than just one.  The probability of rejecting the parabola by accident turns out to be 12.95\%. This sole value is too large to draw any definitive conclusion.

On the other hand, the residuals of the parabolic fit are highly correlated 
(Pearson's $r=0.95$; Fig.~\ref{fig:AbsPlot}c).  In our Monte-Carlo simulation of genuine binaries, such a high correlation between the residuals is obtained for 0.002\% of the systems only.  There is therefore a strong pressure for rejecting the basic two-body model. 

\section{Photometry of the components}

Using our FORS2 observations, we have also estimated the magnitudes of the two components of Luhman 16AB in the $V$, $R$, $I$, and $z$ bands (Table~\ref{tab:col}). {PSF photometry was done with {\tt DAOPHOT}, using the FORS2 standards E5 and LTT4816. They clearly confirm the flux reversal mentioned by \citet{2013arXiv1307.6916B} as the T dwarf becomes the brightest of the two in the $z$-band.

\begin{table}[htbp]
   \centering
   \caption[]{ \label{tab:col}Apparent magnitudes of the components of Luhman 16AB}
  
   \begin{tabular}{@{} cccc @{}} 
   \hline\hline
            Filters    & Luh 16A & Luh 16B & errors\\
  \hline
      $V$      &  23.25 & 24.07  & 0.10 \\
      $R$       & 18.85  & 19.45  & 0.08\\
      $I$ &  15.29  & 15.57 & 0.06\\
      $z$ & 13.83	& 13.76 & 0.02\\
  \hline 
      \end{tabular}
      \tablefoot{Add 3.47 mag to convert these to absolute magnitudes.}
\end{table}

A comparison between our photometry and the most recent models from the Lyon group \citep[BT-Settl; ][]{2011ASPC..448...91A} is shown in online Fig.~\ref{fig:mag}, where we use  the effective temperatures as derived by \citet{2013ApJ...770..124K}. 
To convert to absolute magnitudes we use the distance of $2.02 \pm 0.02$ pc established in this work, while the error bars reflect uncertainties  
in effective temperatures, distance modulus and photometry.
The model isochrones are plotted for 0.1, 1.0 and 5 Gyrs\footnote{Isochrones in Johnson $VRI$ available 
on \url{http://phoenix.ens-lyon.fr/Grids/BT-Settl/}. In case of the non-standard filter $z$, we convolved the model spectra with the 
filter transmission curve, and converted to magnitudes by assuming $z_{Vega}$=0.}. 
Due to their intrinsic faintness in the optical, late L, and T dwarfs have mostly been studied in the near-infrared. Consequently, there are very few works available for a comprehensive comparison of the optical photometry with models, and with our photometry as well. \citet{2002AJ....124.1170D}  published optical $VRI$ photometry and colours for 28 ultracool dwarfs with distances known from parallax measurements. Their sample contains 17 L-dwarfs and three T-dwarf, but only five of them have spectral type L8 or later. The resulting absolute $I$-band magnitude, M$_I$, from their work is $\sim19$ for L8, and $\sim$19.5 for T2, and while in the $R$-band it is $\sim21.7$, in agreement with our results for Luhman 16AB. \citet{2002MNRAS.331..445D} also find M$_I\sim$19 for the spectral type L8. 
The age of Luhman 16AB has been constrained to be less than about 4.5 Gyrs, based on the presence of the lithium line, while from the absence of low surface gravity indicators in the spectra, it is clear that the system is not young (Burgasser et al. 2013). We therefore expect the 1.5 Gyr isochrones to be the most suitable for comparison. The model isochrones seem to overestimate the flux at the L/T transition in $RIz$, and somewhat underestimate it in the $V$.

\section{Discussion}

The FORS2 data, once combined with the data of \citet{Luhman-2013:a} and \citet{Mamajek-2013:a}, yield a significant improvement of the precision of the parallax derived by  \citet{Luhman-2013:a}.  Yet, the FORS2 positions alone indicate that a two-body system is very unlikely whereas an additional companion would explain the observed wobbles. Such a companion must have a mass lower than the brown dwarfs in the system, because i) it would otherwise have been seen in direct imaging or spectroscopy, and ii) a system with two equal mass brown dwarfs would not be detected as there would be no motion of the photocentre.
A 10 M$_{\rm Jup}$ object in orbit of the primary component of Luhman~16AB, which for simplicity we will assume has a mass of 0.05~M$_\odot$, will have a maximum separation on sky of 0.06 to 0.19 arcsec, for an orbital period between 2 months and 1 year, respectively. If we assume such a companion has a negligible luminosity compared to Luhman~16A, it will induce an astrometric signal between 10 and 32 mas, depending on the orbital period. One could think that a more massive companion would lead to a larger astrometric signal,  but this is not necessarily the case, as such companion would now contribute light to the combined system and although the barycentre may change more, the photocentre -- which is what we detect -- would not. Assuming that the luminosity of a BD scales as the square of its mass \citep[e.g.][]{1993RvMP...65..301B}, we find that a 20 M$_{\rm Jup}$ induces a changes of the barycentre between 20 and 60 mas, but a change of the photocentre between 10 and 30 mas. A more massive companion would lead to even smaller motions of the photocentre. On the other hand, a much smaller companion, e.g. 3 M$_{\rm Jup}$, would lead to a maximum motion of 10 mas for a period of 1 year. Although it is still too early to be able to characterise the possible companion, it seems likely that it has a mass between a few and $\sim$30 Jupiter masses, but the latter could then be discovered by adaptive optics, given the expected separation on the sky. At the least, the above discussion shows that the presence of such a companion is compatible with our astrometric signal.  

If this companion would have a planetary mass (below the deuterium burning limit), this would be the first exoplanet around a brown dwarf discovered by astrometry, and possibly the closest exoplanet to the Sun 
(given that the one about $\alpha$ Cen B is still debated; \citealt{Hatzes13}).
Until now, only eight planets are known around brown dwarfs (\url{http://exoplanet.eu}) and they were found  
via microlensing and direct imaging.  Microlensing allows finding a close-in population 
of planets (0.2--0.9 AU; see, for a summary, \citealt{2013arXiv1307.6335H}). These are likely to 
have formed in a protoplanetary disc. They do show larger planet-to-host mass ratio 
than the imaging discovered planets but nothing is known about the physical nature 
of these planetary mass companions due to the large distances of those systems that 
makes the follow-up studies impossible.
On the contrary, the imaging method finds only massive planetary-mass companions at large distance 
(tens or even several hundreds of AU) from their host star. 
These systems resemble more binary sub-stellar systems than planetary systems 
because the planetary mass companions could not have formed from a proto-planetary 
disc but rather during the collapse and fragmentation of a proto-stellar/proto-BD molecular cloud \citep{2006Sci...313.1279J}.
If this tells us that planets {\it do form} around brown dwarfs, the short period range is still unexplored, and Luhman 16AB provides us with a unique opportunity to probe it.
We also note that 
planets inside binary systems are not unheard of and, for example, \citet{2012A&A...542A..92R} identifies 57 exoplanet host stars with stellar companions.  In the last years, imaging campaigns found stellar companions around several dozen exoplanet host stars formerly believed to be single stars (see e.g.
\citealt{2006ApJ...646..523R,2009A&A...494..373M}). Most of
these exoplanet candidates are in the S-type orbit configuration, with the exoplanet surrounding one stellar component of the binary.
As is well known, multiple systems are stable only if they are hierarchical, with period ratios exceeding 10 (see e.g. \citealt{2013arXiv1303.6645C} and references therein). As companion we tentatively detect has likely a period of a few months only (as otherwise, we wouldn't have been able to detect it in our two-month monitoring), while the orbital period of Luhman 16AB is of the order of 25 years \citet{2013ApJ...770..124K}, this is certainly the case here.
Finally, we note that if one of the two brown dwarfs in Luhman 16AB has indeed a substellar companion in close orbit, it should lead to a radial velocity change of the order of 3--5 km/s.  This assumes that the companion's orbit around the brown dwarf is seen edge-on. As our astrometric measurements seem to tentatively indicate that components A and B of Luhman 16AB are in an almost edge-on orbit, this seems a reasonable assumption for the orbit of the third component. We therefore encourage high precision radial velocity monitoring of the system on short time scales.

\begin{acknowledgements}
It is a pleasure to thank Dr. Petro Lazorenko for useful discussions, as well as an anonymous referee for comments that improved the paper.   RK acknowledges support from FONDECYT through grants No 1130140.
\end{acknowledgements}


\Online

\begin{figure}[tbp]
\includegraphics[width=9cm]{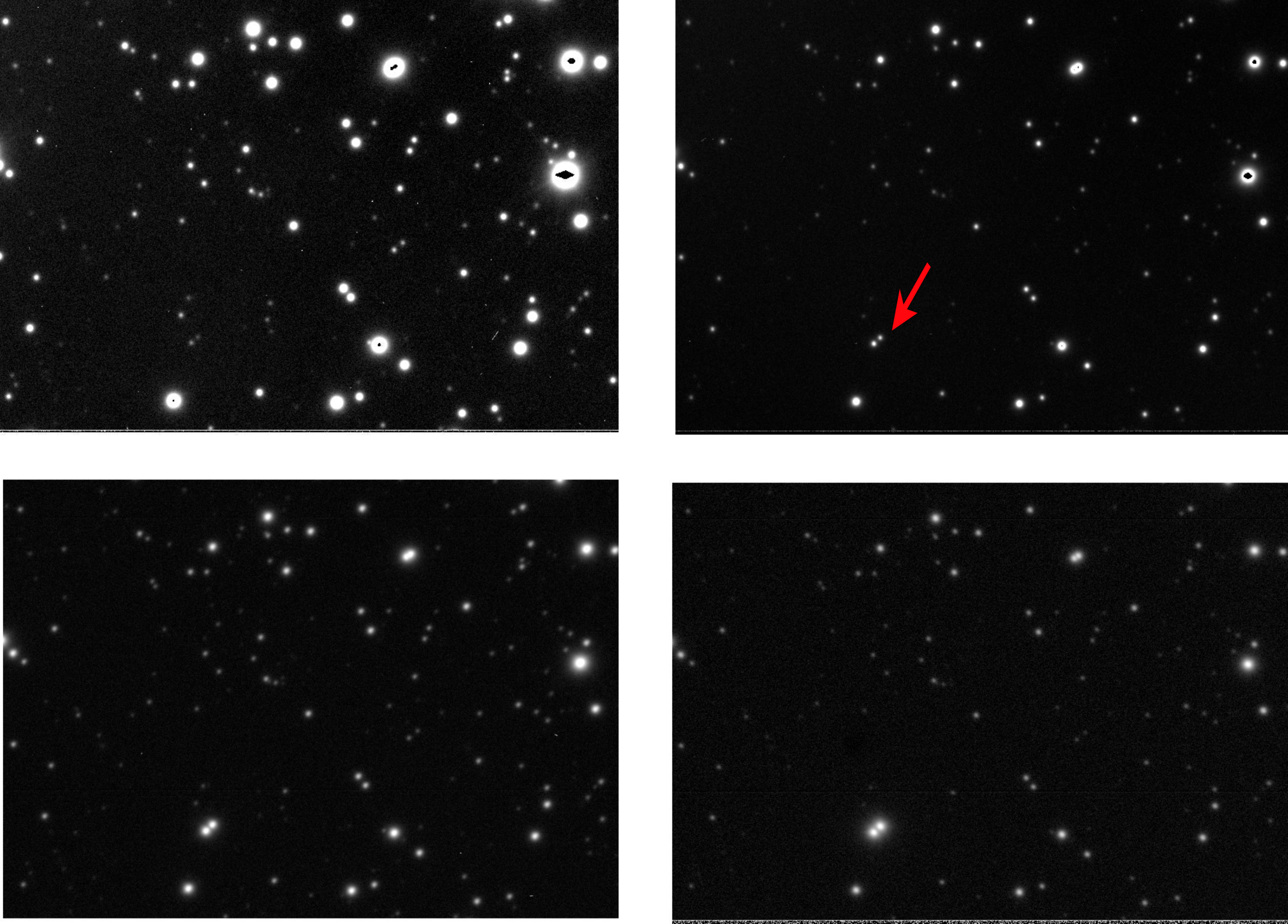}
\caption[]{\label{Fig:Obs}FORS2 observations of Luhman 16AB in various bands: $V$ (top left), $R$ (top right), $I$ (bottom left) and $z$ (bottom right). The two components are indicated in the $R$ band image with an arrow. It is clear from these images that the two components are very red objects. Each image is a very small subset of the full field of view of FORS2, showing only 1.62 $\times$ 1.15 arcmin. North is up and East is to the left.}
\end{figure}

\begin{table*}[htb]
\caption[]{\label{tab:ObsLog}Observation log book.}
\centering
\begin{tabular}{lccllllll}
\hline
\hline
MJD & date & $N$ & $\sigma_{\alpha*}$ (mas) & $\sigma_{\delta}$ (mas) & \multicolumn{2}{c}{Primary} &  \multicolumn{2}{c}{Secondary}\\
& & & & & R.A. COS(DEC) & ~~~~~~~DEC. & R.A. COS(DEC) & ~~~~~~~DEC \\
\hline
56396.99629077 & 2013-04-14 & 18 & 5.174 & 9.913& 96.95709387 &$-$53.31818738 & 96.95681956 &$-$53.31792036\\
56402.0620261  & 2013-04-19 & 17 & 7.570 & 7.619& 96.95708693 &$-$53.31818008 & 96.95680220 &$-$53.31790752\\
56408.04497189 & 2013-04-25 & 18 & 2.704 & 6.539& 96.95705915 &$-$53.31816776 & 96.95677790 &$-$53.31789831\\
56418.03878487 & 2013-05-05 & 16 & 6.515 & 9.875& 96.95702443 &$-$53.31814502 & 96.95675012 &$-$53.31788321\\
56424.09171992 & 2013-05-11 & 18 & 5.187 & 1.392& 96.95701401 &$-$53.31813113 & 96.95673970 &$-$53.31786828\\
56432.98614071 & 2013-05-20 & 36 & 7.548 & 8.585& 96.95698276 &$-$53.31810786 & 96.95671540 &$-$53.31785022\\
56438.03111736 & 2013-05-25 & 18 & 3.955 & 6.190& 96.95698276 &$-$53.31810057 & 96.95670498 &$-$53.31783981\\
56443.99169635 & 2013-05-31 & 18 & 2.262 & 8.045& 96.95696887 &$-$53.31808356 & 96.95668762 &$-$53.31782140\\
56447.99781844 & 2013-06-04 & 18 & 2.175 & 6.385& 96.95695498 &$-$53.31807175 & 96.95668068 &$-$53.31781064\\
56452.98050887 & 2013-06-09 & 36 & 4.466 & 6.947& 96.95694456 &$-$53.31805752 & 96.95666679 &$-$53.31780196\\
56459.97921238 & 2013-06-16 & 13 & 3.917 & 6.894& 96.95693762 &$-$53.31804363 & 96.95666679 &$-$53.31778703\\
56465.97406652 & 2013-06-22 & 22 & 6.364 & 7.023& 96.95693068 &$-$53.31803668 & 96.95665984 &$-$53.31778321\\
\hline
\end{tabular}
\tablefoot{$N$ is the number of images stacked after cropping and the 4th and 5th columns indicate the uncertainty on the positions per epoch (see text),
while the last columns show the mean positions of the two components (in degrees).}
\end{table*}

\begin{figure}[htb]
\includegraphics[width=9cm]{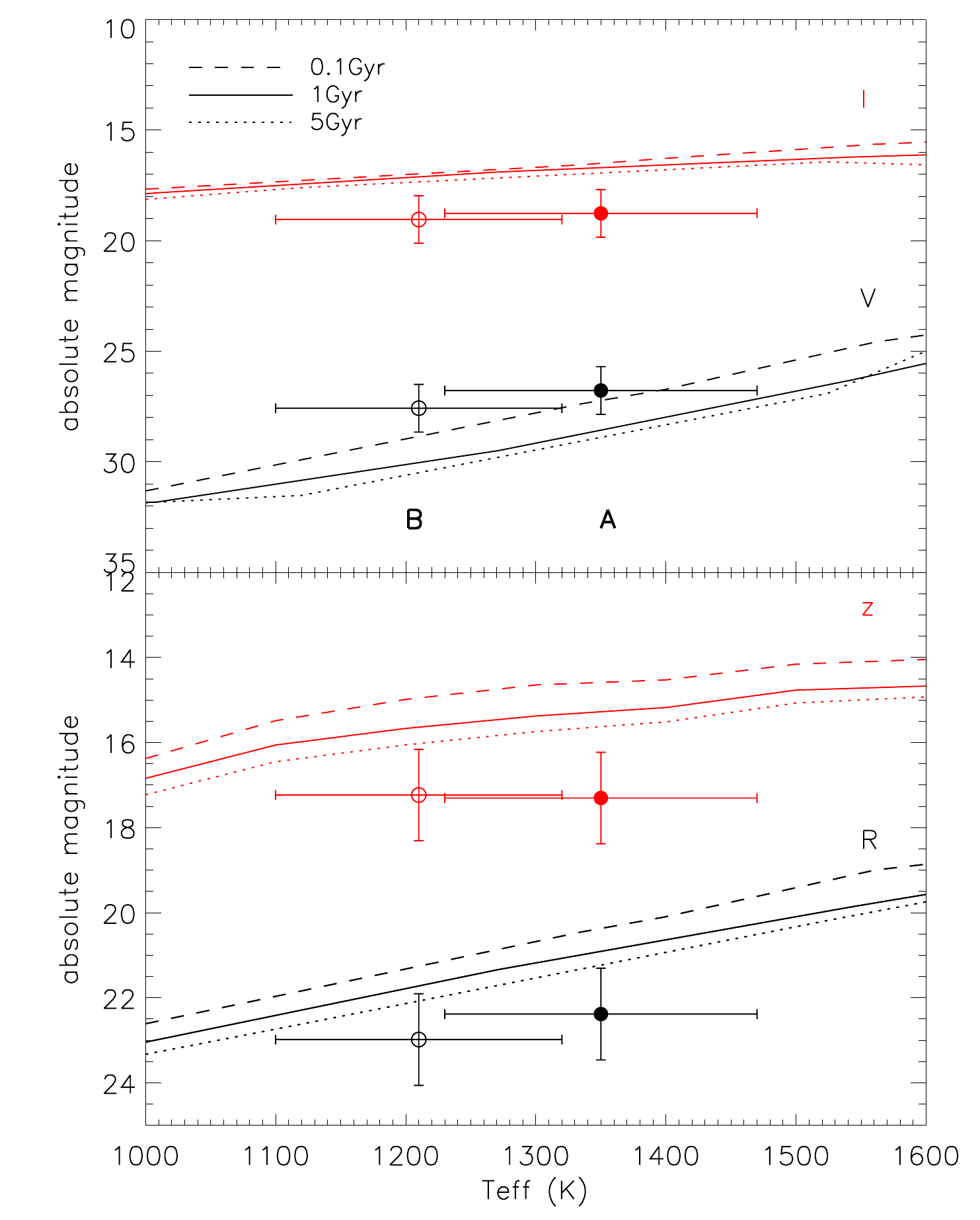}
\caption{Comparisons between observations and models. Circles show the absolute photometry of the system (filled symbols for the primary and open for the secondary), while model isochrones are plotted for 0.1, 1.0 and 5 Gyrs.}
\label{fig:mag}
\end{figure}

\end{document}